# ON THE BEHAVIOR OF HEXANE ON GRAPHITE AT NEAR–MONOLAYER DENSITIES

**Cary L. Pint, M.W. Roth**

Department of Physics, University of Northern Iowa, Cedar Falls, IA  50614-0150

and

**Carlos Wexler**

Department of Physics and Astronomy, University of Missouri-Columbia, Columbia, MO 65211

## ABSTRACT

We present the results of molecular dynamics (MD) studies of hexane physisorbed onto graphite for eight coverages in the range $0.875 \leq \rho \leq 1.05$ (in units of monolayers).  At low temperatures the adsorbate molecules form a uniaxially incommensurate herringbone (UI-HB) solid. At high coverages the solid consists of adsorbate molecules that are primarily rolled on their side perpendicular to the surface of the substrate.  As the coverage is decreased, the amount of molecular rolling diminishes until $\rho = 0.933$ where it disappears (molecules become primarily parallel to the surface).  If the density is decreased enough, vacancies appear.  As the temperature is increased we observe a three-phase regime for $\rho > 0.933$ (with an orientationally ordered *nematic* mesophase), for lower coverages the system melts directly to the disordered (and isotropic) liquid phase.  The solid-nematic transition temperature is very sensitive to coverage whereas the melting temperature is quite insensitive to it, except for at low coverages where increased in-plane space and ultimately vacancies soften the solid phase and lower the melting temperature. Our results signal the importance of molecular rolling and tilting (which result from an the competition between molecule-molecule and molecule-substrate interactions) for the formation of the intermediate phase, while the insensitivity of the system's melting temperature to changing density is understood in terms of in-plane space occupation through rolling. Comparisons and contrasts with experimental results are discussed.

# I. INTRODUCTION

The study of quasi-2D systems (such as atoms and molecules adsorbed onto a surface) has become the focus of much activity over the past few decades. In particular, the adsorption of alkanes has been of interest because of the many applications that these simple hydrocarbons have to commercial lubricants and adhesives. Specifically, hexane ($C_6H_{14}$, or $CH_3$-$(CH_2)_4$-$CH_3$) is a member of the family of straight chained *n*-alkanes whose members differ only in their length ($C_nH_{2n+2}$, or $CH_3$-$(CH_2)_{n-2}$-$CH_3$). This family is simple in structure compared to other organic molecules, yet they still exhibit many internal degrees of freedom thus representing an interesting challenge to study.

When modeling adsorption of molecules such as alkanes, there are a large variety of possible surfaces that could be used to study the phases and dynamics of the adsorbed system. Arguably, graphite is one of the best candidates to implement in such an endeavor as a substrate because it exhibits good mechanical stability, is very readily available and has a high degree of symmetry. As a result of its properties, there has been a wealth of experimental and theoretical work that has been completed on systems involving graphite.[1,2]

The first experimental study of hexane on graphite was conducted by Krim *et al.*[3] using both low-energy electron diffraction (LEED) and neutron diffraction. In this study, for low coverages there is an observed uniaxial incommensurate (UI) herringbone phase below ca. $T = 151$ K, where a first-order melting transition is found. Then, as coverage is increased, the UI phase evolves continuously into a $2 \times 4\sqrt{3}$ commensurate structure at monolayer completion. In more recent experimental work, Taub and co-workers conducted X-ray[4] and neutron[5] diffraction studies of hexane on graphite for submonolayer[4], monolayer[4-7], and multilayer[4] coverages. For a complete monolayer, they observed that the system forms a commensurate herringbone structure at low temperatures, followed by a transition into a rectangular solid/liquid coexistence region by $T = 150$ K, with melting finally occurring at approximately $T = 175$ K[4-7]. At submonolayer coverages, a structure corresponding to a UI phase composed of commensurate herringbone structures separated by low-density fluid filled domain walls was proposed[4].

Other work conducted by Hansen *et al.* has consisted of both computational modeling as well as neutron and X-ray diffraction studies of butane and hexane on graphite at monolayer completion[6,7]. The diffraction studies of hexane on graphite found a loss of translational order at about $T = 150$ K into a phase with short-range order that is thought to involve mobile rectangular-centered (RC) islands within a fluid. Then, at about $T = 170$ K, the system melts into a fluid. Molecular dynamics (MD) simulations of this system show that melting comes about by the coupled effect of large amounts of gauche defects and out-of-plane tilting effectively labeled as the "footprint reduction" mechanism, which allow the necessary creation of in-plane room for the system to undergo melting. However, the temperature of the melting transition is higher than that proposed by experiment, and was soon followed by a string of MD simulations of hexane on graphite by Velasco and Peters[8] which found that a lower adsorbate-substrate interaction gave results that corresponded well to experiment, but also found the effects of gauche defects and molecular tilting to be less dramatic under such conditions.

Another study of monolayer[9] and bilayer[10] hexane was conducted by Peters and coworkers using an MD simulation, but implementing two different methods to model the system, the anisotropic united atom (AUA) model to be compared with the standard united atom (UA) model. For both models, the phases seem to have very good agreement with each other and with experimental work with the only proposed difference being that the isotropic model produces more in-plane mobility from vacancy creation due to increased molecular tilting at all temperatures.

Recent MD simulations of monolayer hexane on graphite found an extended orientationally ordered, "nematic," phase, with a transition from a commensurate herringbone solid to an orientationally ordered liquid crystal at about $T = 138$ K. This phase persists until about $T = 176$K, where the system then melted into an isotropic fluid. In this study, the results of the phase transitions and other calculated quantities (e.g. $S(Q)$ and $g(r)$) seem to be in good qualitative agreement with experiment.[11,12]

Since hexane on graphite is considered as a (quasi) 2-D system (due to the strong graphite-alkane *holding* potential), it is possible that the theory Kosterlitz-Thouless-Halperin-Nelson-Young (KTHNY)[13-17] theory of melting in two dimensions would be applicable to some degree, even though the hexane molecules *are* able to librate out-of-plane, and the substrate corrugation also produces a significant "polarizing" effect. Recent extensive MD simulations of hexane monolayers on graphite have verified that many aspects of the KTHNY theory of melting are seen in these systems[12].

Even though hexane on graphite has been studied very well experimentally, there are still many theoretical questions left unanswered about the phases and dynamics of the system, the investigation of which are imperative to a full understanding of the transitions and phases of hexane on graphite. In addition current simulations suggest that in-plane room has a very strong influence on molecular rolling and tilting, in turn affecting various phase transitions exhibited by the system. Therefore, it is of considerable interest to study this system at coverages somewhat different than the previous extensive studies performed at completion[12].

Specifically, the purpose of this work is to: (i) study the continuous evolution of some of the interesting properties of monolayer hexane on graphite as the density is varied from completion; (ii) gain an understanding of the phase transitions and phases of near-monolayer hexane on graphite; (iii) study the effects that added or reduced in-plane room has; (iv) to further the study of the properties of the nematic (liquid crystal) phase that was observed at monolayers[12], and understand the driving mechanism for the solid-to-nematic phase transition; and (v) to compare the near-monolayer results to hexane at full monolayer coverages[12], thereby placing the complete monolayer results in a broader context.

This paper will be organized in the following way: In Sec. II, the potential model that was used for the simulations are described. Section III describes the details of the numerical simulations. In Sec. IV, the results of the simulation are presented, followed by Sec. V which discusses these results. Finally, our conclusions are presented in Sec. VI.

## II. THE POTENTIAL MODEL

The potential model used in this study consists of both bonded and non-bonded interactions corresponding to interactions between (pseudo) atoms within a molecule, and between atoms of different molecules (or the graphite), respectively.

### A. Non-bonded interactions

The first of the non-bonded interactions is the adsorbate-adsorbate interaction, and is modeled by the well known Lennard Jones pair potential function

$$u_{LJ}(r_{ij}) = 4\varepsilon_{ij}\left[\left(\frac{\sigma_{ij}}{r_{ij}}\right)^{12} - \left(\frac{\sigma_{ij}}{r_{ij}}\right)^{6}\right]. \qquad (1)$$

In Eq. (1), $\varepsilon$ refers to the well depth of the potential, and $\sigma$ represents the collision diameter. Lorentz-Bertholot combining rules:

$$\sigma_{ij} = \frac{\sigma_i + \sigma_j}{2}, \qquad \varepsilon_{ij} = \sqrt{\varepsilon_i \varepsilon_j}, \qquad (2)$$

are used to describe mixed interactions when particles $i$ and $j$ are of different types.

**Table I**. Non-bonded potential parameters used in the simulations.

| Parameter | Value |
|---|---|
| $\varepsilon_{CH3-CH3}$ | 72 K |
| $\sigma_{CH3-CH3}$ | 3.923 Å |
| $\varepsilon_{CH2-CH2}$ | 72 K |
| $\sigma_{CH2-CH2}$ | 3.923 Å |
| $Q$ | 2 |
| $a_s$ | 5.24 Å$^2$ |
| $D$ | 3.357 Å |
| $\varepsilon_{gr}$ | 44.89 K |
| $\sigma_{gr}$ | 3.66 Å |

The other non-bonded potential used is the graphite surface potential, given by a Fourier expansion proposed by Steele[18]:

$$u_i^{gr} = E_{0i}(z_i) + \sum_{n=1}^{\infty} E_{ni}(z_i) f_{ni}(x_i, y_i), \qquad (3)$$

where

$$E_{0i}(z_i) = \frac{2\pi q \varepsilon_{gr} \sigma_{gr}^6}{a_s}\left(\frac{2\sigma_{gr}^6}{45d(z_i+0.72d)^9} + \frac{2\sigma_{gr}^6}{5z_i^{10}} \cdot \frac{1}{z_i^4} - \frac{2z_i^2 + 7z_id + 7d^2}{6d(z_i+d)^5}\right), \qquad (4)$$

$$E_{ni}(z_i) = \frac{2\pi\varepsilon_{gr}\sigma_{gr}^6}{a_s}\left[(\frac{\sigma_{gr}^6}{30})(\frac{g_n}{2z_i})^5 K_5(g_n z_i) - 2(\frac{g_n}{2z_i})^2 K_2(g_n z_i)\right], \quad (5)$$

and

$$f_1(x_i, y_i) = -2\cos\left[\frac{2\pi}{a}(x+\frac{y}{\sqrt{3}})\right] - 2\cos\left[\frac{2\pi}{a}(x-\frac{y}{\sqrt{3}})\right] - 2\cos\left[\frac{4\pi}{a}(\frac{y}{\sqrt{3}})\right]. \quad (6)$$

Here $g_n$ is the modulus of the $n^{th}$ graphite reciprocal lattice vector and the $K$'s are modified Bessel functions of the second kind. The interaction is obtained by summing over an infinite number of graphene sheets using the Euler-MacLaurin Theorem. Only $f_1(x_i, y_i)$ is defined above because the sum in Eq. (3) converges rapidly and only the $n = 1$ term is necessary. All parameters for non-bonded interactions are given in **Table I**.

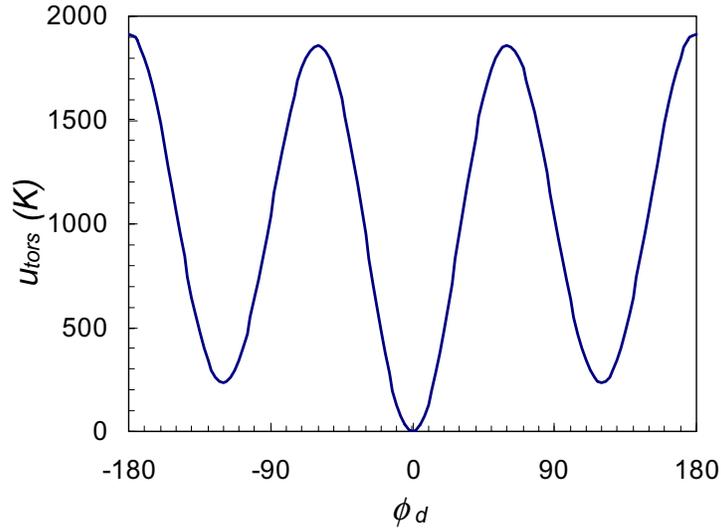

**Figure 1.** Torsional potential $u_{tors}(\phi)$. The absolute minimum corresponds to the *trans* configuration, the two local minima at $\phi = \pm 120°$, $u_{tors} = 234$ K, are the *gauche* configurations.

## B. Bonded Interactions

In this work, there are two bonded interactions that are used, bond angle bending and dihedral angle bending (torsion). All bond lengths are held constant at 1.54 Å with the RATTLE algorithm, which allows for constrained solution of the equations of motion[19]. The first bonded interaction is bond angle bending. Assuming the bond angles to be harmonic, the potential[20] can be expressed as

$$u_{bend} = \frac{1}{2}k_\theta(\theta_b - \theta_0)^2, \quad (7)$$

where $\theta_b$ is the bond angle, $\theta_0$ is the equilibrium bond angle and $k_\theta$ is the angular stiffness. The other bonded interaction is dihedral (torsional) bending, which is of the form[21]

$$u_{tors} = \sum_{i=0}^{5} c_i (\cos \phi_d)^i, \tag{8}$$

where $\phi_d$ is the dihedral angle and the $c_i$ are constants. **Figure 1** shows the torsional potential. All parameters for bonded interactions are given in **Table II**.

**Table II**. Bonded potential parameters used in the simulation.

| Parameter | Value |
|---|---|
| $k_\theta$ | 62793.59 K/rad$^2$ |
| $\theta_0$ | 114º |
| $c_0$ | 1037.76 K |
| $c_1$ | 2426.07 K |
| $c_2$ | 81.64 K |
| $c_3$ | -3129.46 K |
| $c_4$ | -163.28 K |
| $c_5$ | -252.73 K |

## III. SIMULATION DETAILS

All simulations in this study use a constant particle number, planar density, and temperature ($N = 672$, $\rho$, $T$) MD ensemble to atomistically simulate of 112 hexane molecules (each containing 6 pseudo-atoms). In this work, there are eight densities that were studied for hexane, $\rho = 0.875$, 0.903. 0.933, 0.965, 1.00, 1.02, 1.035, and 1.05. For each case, the computational cell size was adjusted in the *y* direction. Periodic boundary conditions (PBC) are utilized in the *x* and *y* directions and free boundary conditions are used in the *z* direction. For all simulations, a velocity Verlet algorithm is used to integrate the equations of motion with a time step of 1 fs. To keep the simulation at constant temperature throughout the simulation, the velocities are rescaled to maintain the center-of-mass, rotational and internal temperatures at the simulated temperature (no significant differences were found when the "thermalization" was performed over a subset of the temperatures below, as this system is good at finding its own *equipartition*):

$$T_{CM} = \frac{1}{3N_m k_B} \sum_{i=1}^{N_m} M_i v_{i,CM}^2,$$

$$T_{ROT} = \frac{1}{3N_m k_B} \sum_{i=1}^{N_m} \boldsymbol{\omega}_i^T \vec{I}_i \boldsymbol{\omega}_i, \qquad (9)$$

$$T_{INT} = \frac{1}{2n_C - 5} \sum_{i=1}^{N_m} \sum_{j=1}^{n_C} m_{ij} (\mathbf{v}_{ij} - \mathbf{v}_{i,CM} - \boldsymbol{\omega}_{i,CM} \times \mathbf{r}_{ij,CM})^2.$$

Here $T_{CM}$, $T_{ROT}$ and $T_{INT}$ are the respective temperatures of the system and the index $i$ runs over molecules and the index $j$ runs over pseudo-atoms within each molecule. All variables indexed with $i$ are standard (velocities $\mathbf{v}$; angular velocities $\boldsymbol{\omega}$, moments of inertia $\vec{I}$) and apply to the $i^{th}$ molecule and those indexed with ($ij$) apply to the $j^{th}$ atom within molecule $i$. Note that since there are both bond angle bending as well as dihedral torsional degrees of freedom, keeping the internal temperature constant gives the system the latitude of partitioning the internal energy.

In all simulations conducted in this study $2 \times 10^5$ steps were used to equilibrate the system, followed by a $5 \times 10^5$ steps to calculate averages and accumulate distributions. The results of these calculations will be presented in the following section.

## IV. RESULTS

For each of the eight average densities that were studied, many runs were carried out in the temperature range of 40-200K to capture the three phases that were observed in previous work on monolayer hexane[12]. This section will be split up into three separate parts. The first part will provide a comparison of order parameters and structural properties of the eight different densities. The second part will analyze the energetics, and the third part will explore various distributions that give insight into the behavior of the system. In many cases results from previous work for the complete hexane monolayer[12] is included for perspective.

### A. Structure and order parameters

For each density above $\rho = 0.933$ there are three distinct phases that are observed. The first phase is the low temperature commensurate solid herringbone phase, followed at higher temperatures by a transition into an orientationally ordered incommensurate nematic (liquid crystal) phase, which is then followed by a transition into an isotropic fluid. Below $\rho = 0.933$ the nematic mesophase is absent. To give a visual appreciation for molecular rolling in the solid, **Figure 2** shows representative herringbone solid phases at various densities. Furthermore, the effect of density on stacking in the nematic mesophase is shown by the representative phases for various densities in **Figure 3**.

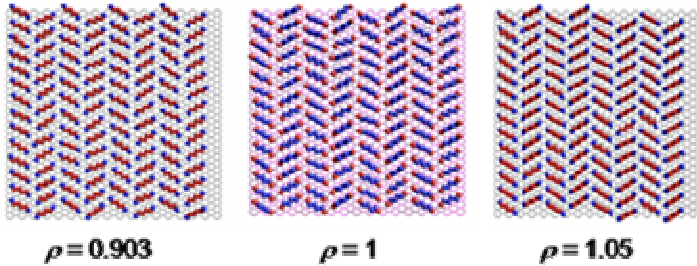

$\rho = 0.903$ $\rho = 1$ $\rho = 1.05$

**Figure 2.** (Color online) Snapshots of herringbone solid phases of hexane on graphite for various densities at $T = 70K$. Note that at $\rho = 0.903$ there is no appreciable rolling of the molecules on their sides while at $\rho = 1.05$ almost all the molecules are rolled. Methyl ($CH_3$) groups are blue and methylene ($CH_2$) pseudoatoms are red.

To effectively describe the structural behavior of the system, several order parameters are utilized. The first order parameter is the herringbone order parameter and is defined as

$$OP_{her} = \frac{1}{N_m} \left\langle \sum_{i=1}^{N_m} (-1)^j \sin(2\phi_i) \right\rangle, \qquad (10)$$

where $\phi_i \in [0, 180°]$ and is the angle that the smallest moment of inertia axis that molecule $i$ makes with the $x$-axis (note that since hexane molecules are not polar, angles are defined in the [0°,180°] range). The integer $j$ is defined to take the difference in orientation of sublattices that are visible in **Figure 2** and **Fig**ure **3**.

From Eq. (10) one finds that $OP_{her}$ takes on a value of unity if all $\phi_i \in \{45°, 135°\}$ and vanishes when $\{\phi_i\}$ is randomly sampling angles in the entire $(x,y)$ plane. All densities studied in the low temperature configuration seem to have a value of $OP_{her} \cong 0.84$, very close the limiting value of the order parameter for a perfectly static herringbone hexane lattice $\phi_i \in \{30°, 150°\}$. This order parameter, along with the following ones are plotted as function of the temperature for various coverages in **Figure 4**.

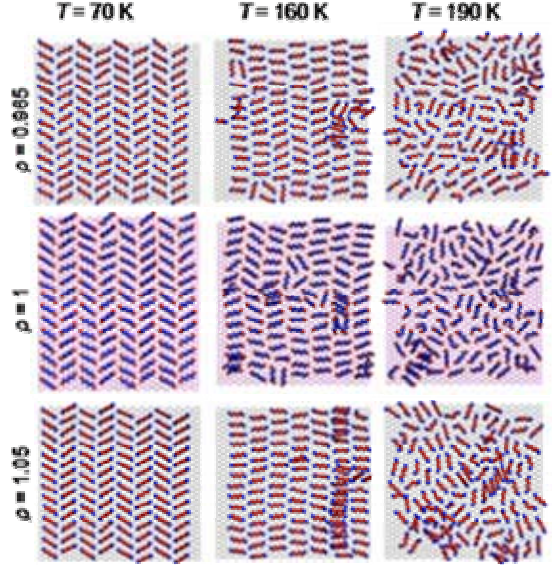

**Figure 3.** (Color online) Snapshot typical configurations of the herringbone solid phase (left), nematic mesophase (middle) and isotropic liquid (right) for $\rho = 0.965$ (top) $\rho = 1$ (middle) and $\rho = 1.05$ (bottom). Note the profound effect of average density on the degree of molecular stacking in the mesophase.

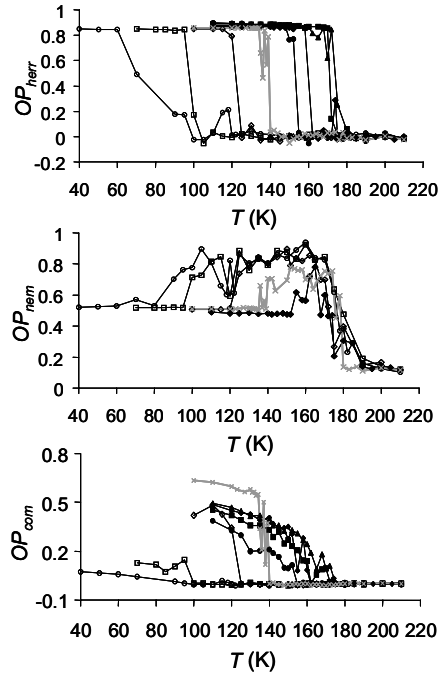

**Figure 4.** $OP_{her}$, $OP_{nem}$, and $OP_{com}$ as functions of temperature for various densities. Solid circles are for a density $\rho = 0.875$, solid squares for 0.903, solid diamonds for 0.933, solid triangles for 0.965, open diamonds for 1.02, open squares for 1.035 and open circles for 1.05. Results for monolayer completion ($\rho = 1$) are included[12] and are represented using points connected with the bold gray line. $OP_{nem}$ curves are shown for only those densities where a nematic phase is exhibited.

Another order parameter that monitored throughout the course of the simulation is the nematic order parameter, defined as

$$OP_{nem} = \frac{1}{N_m} \sum_{i=1}^{N_m} \langle \cos 2(\phi_i - \phi_{dir}) \rangle, \tag{11}$$

where $\phi_i$ is the same angle that is defined for $OP_{her}$ and

$$\phi_{dir} = \frac{1}{2} \tan^{-1} \left[ \frac{\sum_{i=1}^{N_m} \sin(2\phi_i)}{\sum_{i=1}^{N_m} \cos(2\phi_i)} \right], \tag{12}$$

where the 4-quadrant version of $\tan^{-1}$ is used. The nematic order parameter shows a sharp increase when the system undergoes a transition from the herringbone solid to the mesophase.

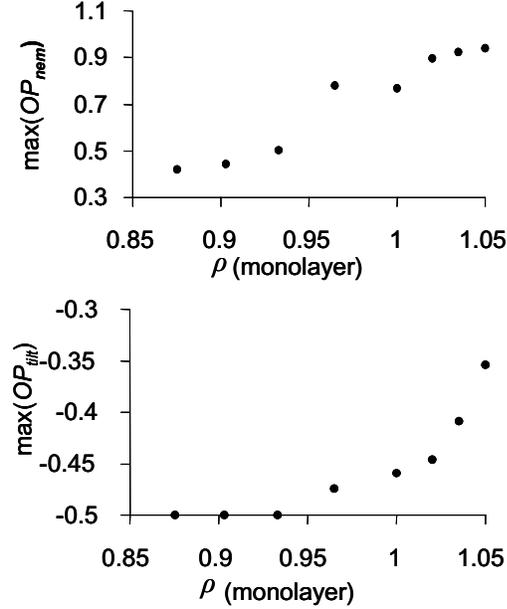

**Figure 5**. Density dependence of the maximum value of $OP_{nem}$ in the nematic phase (top) and maximum value of $OP_{tilt}$ prior to the existence of the nematic phase (bottom).

Another useful order parameter in delineating the adlayer's structure the commensurate order parameter, defined as

$$OP_{com} = \frac{1}{6N_m} \sum_{i=1}^{N_m} \left\langle \sum_{s=1}^{6} \exp(-i\mathbf{g}_s \cdot \mathbf{r}_i) \right\rangle, \tag{13}$$

where the outer sum runs over each of the positions of the molecular centers of mass $\mathbf{r}_i$, and the inner sum runs over all six graphite reciprocal lattice vectors $\mathbf{g}_s$. This order parameter gives valuable information about the position of each molecule over the graphite hexagon centers, and therefore tells whether each molecule is in "registry" with the graphite substrate. If the centers of mass of all molecules are centered over the graphite hexagons, $OP_{com}$ takes on a value of unity. Likewise, if all molecules are randomly sampling positions in the $(x,y)$ plane, then $OP_{com}$ vanishes. Especially since this work deals with incommensurate herringbone solid phases it is essential to use different order parameters to determine the location of phase transitions exhibited by the system.

The last order parameter that we monitored in this work is the tilt order parameter, $OP_{tilt}$, and is defined as:

$$OP_{tilt} = \frac{1}{2N_m} \left\langle \sum_{i=1}^{N_m} (3\cos^2 \theta_i - 1) \right\rangle, \tag{14}$$

where $\theta_i$ is the angle that the smallest moment of inertia axis of molecule $i$ makes with the vertical. $OP_{tilt}$ is the thermal average of a Legendre polynomial ($P_2$), and takes a value of $-0.5$ if the long axis of each hexane molecule is parallel to the $(x,y)$ plane. **Figure 5** shows a plot of the maximum value $OP_{nem}$ assumes and the maximum value $OP_{tilt}$ assumes prior to melting for various densities. The molecular tilting behavior of the system will be described in detail later.

## B. Energetics

Along with order parameters and other structural indicators, it is important to understand the energetics of the system In this study, two quantities very descriptive of system behavior are the average Lennard-Jones energy, $<U_{LJ}>$, and the average corrugation potential energy, $<U_1>$. The thermal average of the Lennard-Jones energy per molecule is defined as:

$$<U_{LJ}> = \frac{1}{N_m} \left\langle \sum_{i=1}^{N} \sum_{j=i+1}^{N} u_{LJ}(r_{ij}) \right\rangle, \tag{15}$$

and is a very useful quantity in targeting the commensurate-incommensurate (CIT) phase transition where the molecule-molecule interactions dominate over the molecule-substrate interactions. Likewise, the average of the lateral Steele *corrugation energy* per molecule is defined as:

$$<U_1> = \frac{1}{N_m} \left\langle \sum_{i=1}^{N} E_{1i}(\vec{r}_i) \right\rangle, \tag{16}$$

and is related to $OP_{com}$ defined in the previous subsection with only the difference that it gives information about the atomic order of the substrate. Similar to $OP_{com}$, if the molecules are randomly sampling positions in the $(x,y)$ plane, then $<U_1>$ will take on the smallest value. Unlike $OP_{com}$, however, the corrugation energy is calculated by summing over all pseudoatoms in the adsorbate molecules, which is a very important difference that will be discussed later. The temperature dependence of both $<U_1>$ and $<U_{LJ}>$ as functions of temperature for all densities studied are shown in **Figure 6**.

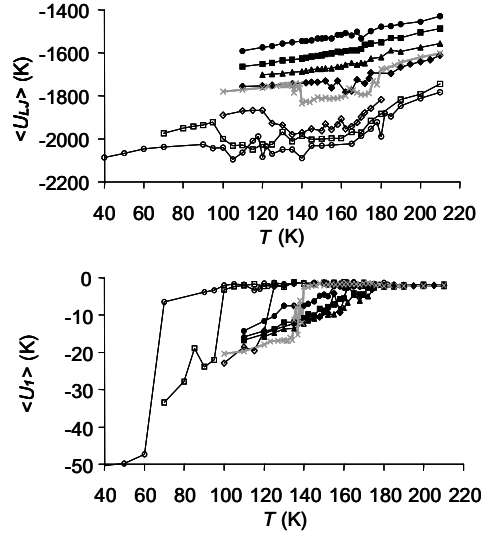

**Figure 6.** Temperature dependence of the average Lennard-Jones interaction $<U_{LJ}>$ and the average corrugation energy $<U_1>$ per molecule. Format is the same as in **Figure 4**.

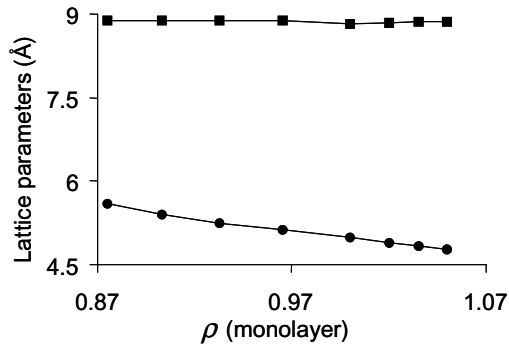

**Figure 7**. Lattice parameters $a$ (top, squares) and $b$ (bottom, circles) for the low-temperature solid at various densities.

**Figure 7** shows lattice parameters (*a,b*) for the system as determined from the pair correlation function $g(r)$ [see Ref. 12 Sec. III.E for a description of the calculation of $g(r)$]. The lattice parameters are useful in determining the structure of the solid phase and in what direction the system is incommensurate. Moreover, **Table III** and **Figure 8** show the values for the herringbone-to-nematic phase transition $T_1$ and the nematic to isotropic fluid (melting) transition, $T_2$ as determined by locating the temperatures at which order parameters and various energies exhibit the greatest rate of change with temperature. The nematic order parameter is unique in that it can locate both phase transitions. The temperature of the various phase transitions are also substantiated by the change in character of intermolecular pair correlation functions.

**Table III**. Temperatures for the herringbone-to-nematic phase transition $T_1$ and the nematic to isotropic fluid (melting) transition, $T_2$ determined from the behavior of $OP_{her}$, $OP_{com}$ as well as system energetics.

| **Density (monolayers)** | $T_1$ **(K)** | $T_2$ **(K)** |
|:---:|:---:|:---:|
| 0.875 | N/O | $155 \pm 3K$ |
| 0.903 | N/O | $172 \pm 3K$ |
| 0.933 | N/O | $174 \pm 3K$ |
| 0.965 | $155 \pm 5K$ | $175 \pm 3K$ |
| 1 | $138 \pm 2K$ | $176 \pm 3K$ |
| 1.02 | $122 \pm 3K$ | $172 \pm 3K$ |
| 1.035 | $98 \pm 3K$ | $175 \pm 3K$ |
| 1.05 | $85 \pm 3K$ | $175 \pm 3K$ |

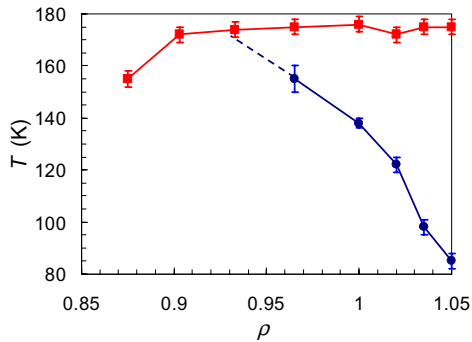

**Figure 8.** (Color online) Temperatures for the herringbone-to-nematic phase transition $T_1$ (blue circles) and the nematic to isotropic fluid (melting) transition $T_2$ (red squares) from **Table III**. Note that $T_1$ is *projected* to be close to $T_2$ at ca. 0.933 monolayer coverage, at which point the intermediate nematic phase is no longer observed.

## C. Distributions

The first distribution that is shown in **Figure 9** is the molecular bond roll angle distribution, $P(\psi)$, with $\psi$ being defined for a bond consisting of three consecutive pseudo-atoms as:

$$\psi = \cos^{-1}\left\{\frac{\left[(\vec{r}_{j+1}-\vec{r}_j)\times(\vec{r}_{j-1}-\vec{r}_j)\right]\cdot\hat{z}}{\left|(\vec{r}_{j+1}-\vec{r}_j)\times(\vec{r}_{j-1}-\vec{r}_j)\right|}\right\}. \tag{17}$$

This roll-angle takes on a value of 0° when the plane consisting of the three molecules in the bond is parallel to the graphite substrate, and takes on a value of 90° when the plane is perpendicular to the substrate. The roll-angle is useful in illustrating to what degree the molecules are rolled on their sides at various densities. **Figure 9** shows bond-roll distributions throughout the entire temperature range for all eight densities examined. Another distribution which can place bond rolling in broader perspective is the atomic height distribution, $P(z)$, shown in **Figure 10**. $P(z)$ not only can show molecular rolling behavior but can also show signatures of changes in other vertical behavior of the adsorbate such as tilting and stacking.

# V. DISCUSSION

This section will be split up into 3 sub-sections. The first section will discuss the solid herringbone phase, concentrating on the herringbone-nematic phase transition and will explore the mechanism for this transition. The second section will discuss aspects of the nematic phase that is observed as the density is increased, and the last section will discuss the nematic-to-isotropic fluid phase transition and analyze how this transition takes place.

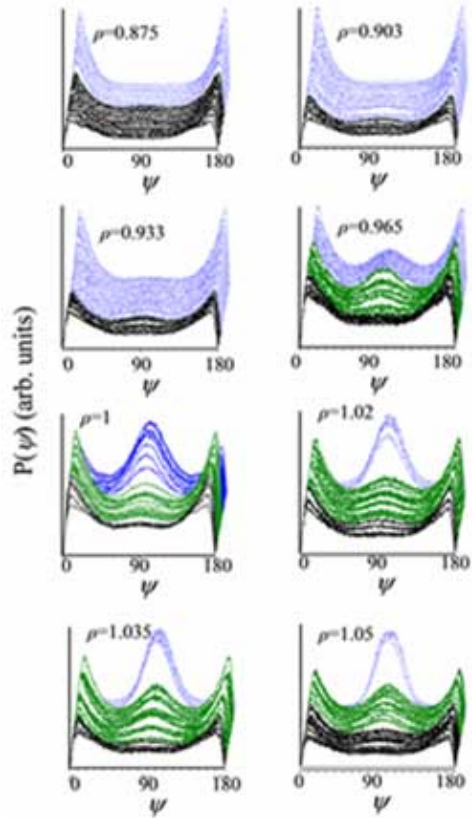

**Figure 9.** (Color online) Bond roll angle distributions, $P(\Psi)$, for various temperatures and at all densities examined in this study. The regular blue lines correspond to the low temperature herringbone phase, the medium olive lines correspond to points in the nematic phase, and the bold black lines correspond to the isotropic fluid.

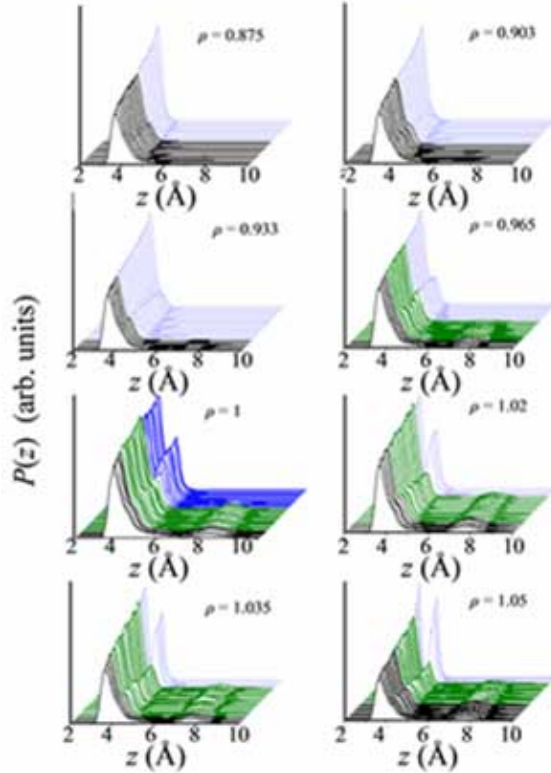

**Figure 10.** (Color online) Atomic height distributions, $P(z)$ for various temperatures at all the densities examined ranging from $\rho = 0.875$ (upper left) to $\rho = 1.05$ (lower right), including $\rho = 1$. Color scheme is the same as in **Figure 9**.

### A. Herringbone Solid-to-Nematic Transition

As shown in **Table III** and **Figure 8** the solid-nematic transition temperature $T_1$ depends strongly on density. Below $T_1$, where the system is in a herringbone solid configuration whose commensurability depends strongly on the system's average density. To begin with, the behavior of $OP_{com}$ at low temperature shows that the system is more incommensurate as the density departs from $\rho = 1$, and the lattice parameters in **Figure 7** reveal that the system remains commensurate in the $x$ direction but is uniaxially incommensurate in the $y$ direction. Seemingly in contradiction to the information from $OP_{com}$ in **Figure 4**, examination of **Figure 6** shows that the corrugation potential energy is increasingly stronger as the density is increased over the entire density range. Such a result is understood by the fact that $OP_{com}$ is a center of mass quantity which shows that the system is incommensurate but $<U_1>$ is an atomistic quantity, which illustrates that even when the system is incommensurate the pseudoatoms are able to reorient so as to better fit in the graphite potential wells. Also, despite the drastic change in commensurability of the system between the eight densities examined, the magnitude of $OP_{her}$ in **Figure 4** exhibits no significant change in its low-temperature value of around 0.84. Such a results means that the values of $\phi_i$ in Eq. (10) remain consistent regardless of the commensurability and the low-temperature solid structure is always herringbone.

The pair correlation functions (not shown here) for the eight densities in the herringbone phase are very different. One of the significant differences is the positioning of various peaks which shows up in the behavior of the calculated lattice parameters (*a*,*b*) (**Figure 7**). It also seems that the molecular neighbors are less defined for higher densities than for lower densities. At lower densities we find that the molecular neighbors are easily identifiable by considering each neighbor to span from one local minima over the intermolecular portion of $g(r)$ to the next. This task is virtually impossible for $\rho = 1.05$, except for possibly the peak corresponding to the second molecular neighbors, which seems to be fairly distinct. This behavior of $g(r)$ at higher densities could be partly due to the greater incommensurability of the system at these densities, as was discussed in the previous paragraph.

Another very interesting aspect of the low-temperature solid herringbone phase is illustrated in **Figure 9** and **Figure 10**, where we find that as the density is increased, the molecules tend to roll perpendicular to the plane at low temperatures. The signature of collective rolling is a peak at $90^0$ in $P(\psi)$ and a double peak in $P(z)$. This means that the lack of in-plane room forces the molecules to roll on their side. Such rolling exhibits a very interesting energy competition. Although it is not energetically favorable with respect to the strong holding potential, **Figure 6** clearly shows that as density is increased in the solid phase the average Lennard-Jones interaction energy $<U_{LJ}>$ decreases dramatically, as does the corrugation energy $<U_1>$ to a much smaller extent. In **Figure 10**, at $\rho = 1.05$, we see that the amplitudes of the two peaks in the low temperature range are *almost* identical, which means that almost all of the molecules in this low temperature solid phase are rolled perpendicular to the graphite surface. Likewise, **Figure 9** and **Figure 10** show that for densities less than or equal to 0.933 there is virtually no rolling and the molecules are mostly flat. This can also be seen in the snapshots in **Figure 2** and **Fig**ure 3. The submonolayer regime has been experimentally examined for $0.6 \leq \rho < 1$[5]. The results of neutron[5] and x-ray[5] diffraction experiments suggest that the submonolayer phase is uniaxially incommen-surate, as our simulations do. However the diffraction patterns are consistent with a system that is commensurate on a short length scale and incommensurate on a longer one, so a striped domain wall structure is proposed[5]. Moreover, at all densities examined, diffraction results[5] show that the molecules do not exhibit the molecular rolling seen in simulations[9-12]. We suspect that such a discrepancy could largely be attributed to the lack of explicit hydrogens on the hexane molecules, which would affect the molecule-substrate interaction greatly. The supermonolayer density regime examined here has not been investigated experimentally and therefore is mainly a tool in this work with which to establish trends needed do draw conclusions about the dynamics of the system. Moreover, we suspect that it gives a reasonable prediction of the melting temperature in this regime.

Inspecting **Figure 5** we see that the maximum tilt order parameter remains at around –0.5 until $\rho = 9.965$ where there is considerable tilting. Moreover, the increase in tilting correlates with increased strength in the nematic phase due to the behavior of the maximum value of $OP_{nem}$. It is important to note that the maximum value of $OP_{tilt}$ occurs at the onset of the nematic phase in all cases. We also find at that the molecular bond roll angle distribution has significant changes at the solid to nematic transition, with many of the molecules that were previously rolled perpendicular to the graphite surface, now relax into a position parallel to the substrate, which is shown by the large amplitude at $\Psi = 0°$ and $180°$. Further, at this transition, from **Figure 10**, that the atomic height distributions show significant number of molecules that are promoted to the second layer via stacking. We find that this is compelling evidence that the driving mechanism for the nematic phase transition is out-of-plane tilting. When all the molecules are rolled perpendicular to the surface, the extra thermal energy needed for the molecules to librate out of the plane is less than the case where some of the molecules are flat in the plane. This also suggests that out-of-plane tilting is directly a result of the molecules rolling, so the competition between their minimizing the holding potential energy by being flat and the minimization of the Lennard-Jones interaction is very important here. This seems to be consistent with the herringbone-nematic transition temperatures that were observed. In addition comparison of $<U_{LJ}>$ in **Figure 6** with $OP_{nem}$ in **Figure 4** clearly shows that the extent of the nematic phase is directly correlated to the drop in Lennard-Jones energy, underscoring the importance of rolling/tilting in the solid to nematic transition in this system because, as discussed before, $<U_{LJ}>$ is minimized at higher densities with the molecules rolled on their sides. At lower densities, **Figure 9** and **Figure 10** show that less molecules are rolled perpendicular to the substrate, and the transition temperature is at a higher value than that observed at $\rho = 1.05$, and is even absent below $\rho = 1.05$.

## B. The Nematic Mesophase

There are some interesting and unique aspects of the nematic phase. Recent work on monolayer hexane originally found a nematic phase[9,10] which was later characterized thermodynamically as a liquid crystal.[12] $OP_{nem}$ in **Figure 4**, $<U_{LJ}>$ in **Figure 6** and the maximum values of $OP_{nem}$ shown in **Figure 5** all show the dependence of the extent of the nematic phase on the system's average density. Clearly the nematic phase increases in intensity and extent with increasing density, suggesting that the Lennard-Jones interaction drives the solid-nematic transition. As the system's average density is increased, although there is a significant number of molecules promoted to the second layer, the molecules that occupy the first layer seem to have a more defined orientational ordering about a common director. This could be another result of the molecules having less in-plane room, leaving the molecules on the first layer to fill the vacancies that are created from molecules being promoted to the second layer. But, unlike monolayer hexane, rather than occupying that space immediately, the molecules will first lower their energy by rolling parallel to the graphite surface, which then leaves them without the mobility to effectively fill that vacancy in such a way that they oppose the director. In monolayer hexane, we found that patches of molecules in the nematic phase are oriented perpendicular to the director angle, which arises from the molecules filling the vacancies that are created by layer promotion in the nematic phase[12]. In this case, we find that there is a more well defined director because of the molecular rolling perpendicular to the substrate, which does not induce molecular mobility to fill the vacancies in the first layer, but rather molecular *rolling* into the plane.

Examining the temperature dependence of the maximum value of $OP_{tilt}$ prior to melting (**Figure 5**) against the backdrop of the other quantities discussed above it is clear that as density is increased the molecules show more tilting at the onset of the nematic phase. Examination of $O_{tilt}$ vs. temperature at various densities (not shown) reveal that there is considerable tilting prior to the formation of the nematic and that for all but the three highest densities, tilt fluctuations decrease after the nematic changes into the isotropic liquid. For the lowest three densities examined there is virtually no tilting and the nematic phase is absent. A picture emerges where, in the solid, molecules that would otherwise lay flat are rolled on their side due to compression at higher densities. Since the rolling minimizes the Lennard-Jones interaction energy between adsorbate molecules but laying flat minimizes the molecule-substrate energy, tilting is the mechanism with which the nematic phase is formed, thus lowering the Lennard-Jones interaction energy and causing the molecules to lay flat and not roll. The rolling behavior also shows up in **Figure 10**, where $P(z)$ has two distinct peaks when the molecules are rolled on their sides. In addition $P(z)$ shows that at the liquid crystal transition, there are a significant number of molecules that are promoted to the second layer in the nematic phase, and this promotion continues throughout the isotropic fluid phase as well.

Our results should be understood with an important caveat. Although the physics involved in our simulations seems reasonable, the corresponding real system more than likely is comprised of nematic islands embedded within a fluid. Although we feel that it would be very useful to conduct simulations of much larger systems to see if this is indeed the case, they would have to be sufficiently large so as to prevent generation of enough data to see trends and make comparisons.

### C. The Nematic-Isotropic Liquid Transition

As **Table III** and **Figure 8** show, the nematic-fluid phase transition takes place at about $T_2 = 172$–$175K \pm 3K$ for all densities studied that exhibit a nematic. As opposed to the solid-nematic transition, we find that this transition is insensitive to coverage until the molecules can no longer fill vacancies by rolling flat, as shown in **Figure 2** and **Figure 3**. One reason for this is that the molecules are able to occupy space by rolling as density decreases but once the density decreased below the point at which they are flat then vacancies form and the melting temperature drops. Previous work on this system found that both gauche defects and out-of plane tilting contribute to the melting transition in monolayer hexane on graphite.[12] We find that our results for supermonolayer hexane are consistent with this idea as evidenced by both the behavior of $OP_{tilt}$ in **Figure 4** and a typical dihedral angle distribution, shown in **Figure 11**. In the behavior of $OP_{tilt}$ with changing temperature (not shown) we find that just the before the melting transition in each case where a nematic is present, there is distinct single peak in the region of the melting temperature that corresponds to the "tilt-lock" mechanism that was found in monolayer hexane[12]. In addition the three highest densities support considerable tilting even in the fluid.

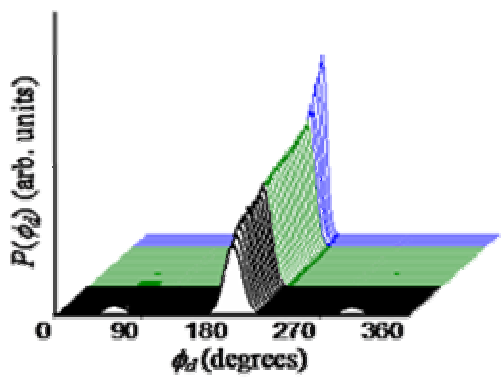

**Figure 11**. (Color online) The dihedral angle distribution $P(\phi_d)$ for $\rho = 1.05$, typical for the systems studied here. Color scheme is the same as in **Figure 9**. Note the proliferation of defects after melting.

We also find the presence of gauche defects become significant after melting, which is also a good indicator that these contribute to the melting transition. In addition when the dihedral potential barrier is made considerably stronger, the melting temperatures for all densities increase by ca. 25 K, indicating again that gauche defect formation is relevant to the melting mechanism. Moreover, the insensitivity of the change in melting temperature change across various densities suggests that the effect is strictly a molecule-substrate one, not involving molecule-molecule coupling. **Figure 9** shows that in the low temperature configurations for higher densities, as the density of molecules on the surface is slightly increased, more molecules tend to roll perpendicular to the substrate. At the transition from the solid to the liquid crystal, there is a dramatic sudden change in the rolling behavior of the molecules as the majority of them relax to a position with their planes parallel to the substrate. We see then at the isotropic fluid transition, there is another change where the small peak at 90° corresponding to molecules being rolled with their planes perpendicular to the surface, disappears. The signatures of the various changes in rolling behavior are also apparent in $P(z)$ in **Figure 10**, which also shows that in the fluid there is a ponderable number of molecules that are promoted to the second layer.

Our simulated melting results are in reasonable agreement with experiment,[5] where $T_2$ is at a little under 170 K at $\rho = 0.79$ and increases slightly to a little over 172 K at monolayer completion. The experimental melting temperature drops to around 152 K at $\rho = 0.6$, which is at a density somewhat higher than seen in our simulations, where the melting temperature drops sharply at a density of 0.875. As before, we suspect that the discrepancy is partly due to the un-derestimation of molecule-surface interactions without explicit hydrogens on the simulated hexane molecules. Were the simulated molecules flat on the surface they would manipulate in-plane space differently, which would alter the melting temperature most at submonolayer densities. The system melts at around $T_2 = 190$ K for the bilayer system ($\rho = 2$) which we do not examine.

## VI. CONCLUSIONS

With this study, we are able to make some important conclusions about hexane on graphite at near-monolayer densities:

*(i)* The transition mechanism from the commensurate herringbone solid to the incommensurate nematic phase involves out-of-plane tilting.

*(ii)* The solid to nematic transition temperature $T_1$ is sensitive to coverage.

*(iii)* The tilting is energetically favorable for the Lennard-Jones hexane-hexane interaction but laying flat is favorable for the molecule-substrate interaction so the nematic is driven by the resulting energy competition. For systems whose densities are too low to allow tilting the nematic phase is absent.

*(iv)* The melting temperature $T_2$ is insensitive to coverage unless the molecules can no longer fill vacancies by rolling flat.

*(v)* At high density the adsorbate molecules are rolled on their sides and as density decreases they flatten, which occupies space, prevents vacancy formation and holds the melting temperature fairly steady. When vacancies form at low enough densities where significant rolling does not occur then the melting temperature drops.

*(vi)* The results of the simulations presented here must be understood within the context that, most likely, after the herringbone solid phase there are nematic *islands* embedded in a fluid. Conducting simulations with very large systems would prove insightful but are also preventative for getting a enough data to see trends.

*(vii)* The low–temperature solid phase for the hexane on graphite system at near – monolayer coverages is a uniaxially incommensurate solid.

*(viii)* Although the supermonolayer densities simulated here have not been experimentally examined they provide an invaluable tool in establishing trends in the system's behavior and could serve in a predictive capacity within the context of the simulation limitations discussed here.

*(ix)* Overall the simulation results are in good agreement with experiment and are felt to be of reasonable reliability. The discrepancies between simulation and experiment can be partially addressed by using explicit hydrogens on the hexane molecules. Such a change would more accurately represent the molecule-substrate interaction and cause the adsorbate molecules to occupy space differently, which would affect the system's structure and dynamics most at submonolayer densities.

## ACKNOWLEDGEMENTS

The authors would like to thank Haskell Taub, Flemming Hansen, and Güenther Peters for useful discussions. Acknowledgment is made to the Donors of the Petroleum Research Fund, administered by the American Chemical Society, for support of this research. CW acknowledges support by the University of Missouri Research Board and the University of Missouri Research Council. CP and MR would like to acknowledge Paul Gray and the UNI computer science department for their helpfulness and computer resources.